\documentclass[prd,floats,floatfix,nofootinbib,superscriptaddress,showpacs,10pt]{revtex4-2}
\usepackage{graphicx}
\usepackage{graphics}
\usepackage[dvips]{color}
\usepackage[dvips]{color}
\usepackage{xcolor}

\usepackage{float}

\newcommand{\be}{\begin{equation}}
\newcommand{\en}{\end{equation}}
\newcommand{\bea}{\begin{eqnarray}}
\newcommand{\ena}{\end{eqnarray}}

\begin{document}

\title{\Large On methods for radiometric surveying in radiotherapy bunkers.} 

\author{Eduardo Sergio Santini}
\email{sergio.santini@cnen.gov.br}
\email{santini@cbpf.br}
\affiliation{Comiss\~ao Nacional de Energia Nuclear - CNEN \\ Coordena\c{c}\~ao Geral de Instala\c{c}\~oes M\'edicas e Industriais - CGMI \\
Rua General Severiano 90, Botafogo, 22290-901,  Rio de Janeiro, Brazil}
\affiliation{Centro Brasileiro de
Pesquisas F\'{\i}sicas - CBPF - COSMO, Rua Xavier Sigaud, 150, Urca,
22290-180, Rio de Janeiro, Brazil}

\author{Renato Vasconcellos de Oliveira}
\email{renato.oliveira@cnen.gov.br}
\affiliation{Comiss\~ao Nacional de Energia Nuclear - CNEN \\ Coordena\c{c}\~ao Geral de Instala\c{c}\~oes M\'edicas e Industriais - CGMI \\
Rua General Severiano 90, Botafogo, 22290-901,  Rio de Janeiro, Brazil}

\author{Nozimar do Couto}
\email{nozimar.couto@cnen.gov.br}
\affiliation{Comiss\~ao Nacional de Energia Nuclear - CNEN \\ Coordena\c{c}\~ao Geral de Instala\c{c}\~oes M\'edicas e Industriais - CGMI \\
Rua General Severiano 90, Botafogo, 22290-901,  Rio de Janeiro, Brazil}

\author{Camila Salata}
\email{camila.salata@cnen.gov.br}
\affiliation{Comiss\~ao Nacional de Energia Nuclear - CNEN \\ Coordena\c{c}\~ao Geral de Instala\c{c}\~oes M\'edicas e Industriais - CGMI \\
Rua General Severiano 90, Botafogo, 22290-901,  Rio de Janeiro, Brazil}

\author{Paulo Antônio Pereira Leal}
\email{paulo.leal@cnen.gov.br}
\affiliation{Comiss\~ao Nacional de Energia Nuclear - CNEN \\ Coordena\c{c}\~ao Geral de Instala\c{c}\~oes M\'edicas e Industriais - CGMI \\
Rua General Severiano 90, Botafogo, 22290-901,  Rio de Janeiro, Brazil}

\author{Fl\'avia Cristina da Silva Teixeira }
\email{flavia.teixeira@cnen.gov.br}
\affiliation{Comiss\~ao Nacional de Energia Nuclear - CNEN \\ Coordena\c{c}\~ao Geral de Instala\c{c}\~oes M\'edicas e Industriais - CGMI \\
Rua General Severiano 90, Botafogo, 22290-901,  Rio de Janeiro, Brazil}

\author{Georgia Santos Joana}
\email{georgia.joana@cnen.gov.br}
\affiliation{Comiss\~ao Nacional de Energia Nuclear - CNEN \\ Coordena\c{c}\~ao Geral de Instala\c{c}\~oes M\'edicas e Industriais - CGMI \\
Rua General Severiano 90, Botafogo, 22290-901,  Rio de Janeiro, Brazil}


\begin{abstract}\label{int}

Radiometric surveys in radiotherapy bunkers have been carried out in Brazil for many years, both by the same radiotherapy facility for verification of shielding as by the regulatory agency for licensing and control purposes.
In recent years, the Intensity Modulated Radiation Therapy (IMRT) technique has been gradually  incorporated into many facilities. Therefore, it has been necessary to consider the increased leakage component that has an important impact on the secondary walls. For that, a radiometric survey method has been used that considers an increased "time of beam - on" for the secondary walls. In this work we discuss two methods of doing this: the first considers that this "time of beam - on" affects the sum of the two components, leakage and scattered. In another method it is considered that only the leakage component is affected by this extended "time of beam - on ". We compare the methods and show that for secondary walls with $U=1$ the first method overestimates dose rates by important percentages and for secondary walls with $U<1$ it can both overestimate or underestimate the dose rates, depending on the parameters of the project. An optimized procedure is proposed, according to the use factor ($U$) of the secondary wall to be measured.
\end{abstract}

\pacs{Radiological Protection, Radiotherapy shielding}
\maketitle


\section{Introduction} 

The licensing process of a linear accelerator (LINAC) radiotherapy facility involves, in the final part of it, a crucial step that consists of a regulatory inspection during which, among many other checks, a radiometric survey is performed on the surroundings of the treatment bunker. In Brazil this has been carried out since several decades by Brazilian Nuclear Regulatory Authority (CNEN). The methodology used in this radiometric survey is to measure the instantaneous dose-equivalent rate in Sv/h at strategic points and calculate the weekly rate through a simple equation that involves the weekly workload $W$ (Eq. (\ref{semIMRT}) below, see for example \cite{paiva}). With the emergence of IMRT technology, the secondary walls began to have to oversee a larger leakage-radiation workload, due to the largest number of monitor units required \cite{NCRP151} \cite{kairn}. This new leakage-radiation workload was generally called $W_L$. So that the radiometric surveys in services that have an IMRT technique began to be performed using, for the secondary walls, the same formula but with the new $W_L$ leakage-radiation workload instead of $W$ 
(i.e. Eq. \ref{velhoW}).
In this work we show that this formula is not completely suitable to determine the weekly dose-equivalent rate $ R $ for a secondary wall. We propose a new method through  a new formula obtained from the equations used in the calculation of shielding, which should replace the historically used. We show that the old formula overestimates dose rates always in cases where the secondary wall has a unit use factor ($ U = 1 $). In cases where the use factor is not equal to 1 we show that the old method can both overestimate and underestimate. 
We experimentally verified these results through varied radiometric surveys that we performed in various radiotherapy services. From these verifications we propose a simple protocol to optimize the radiometric surveys that must be performed as the final stage of the licensing process of the radiotherapy service. This article is organized as follows: after this introduction, in section II we recall the old method then we deduce the correct equation which defines the new method and compare them. In section III we use real results obtained for several bunkers to verify the equations obtained. Section IV is for discussion of our results and proposal of a protocol and section V is for our conclusions.

\section{Method}\label{meth}

We consider a secondary barrier of a LINAC bunker, in a radiotherapy facility that uses IMRT technique.
The  total instantaneous dose-equivalent rate $I_T \,\,  [\frac{Sv}{h}]$,  with the machine operating at the absorbed-dose output rate $\dot{D_o}$ at $1\, m$ in $ [\frac{Gy}{h}]$, measured $30\, cm$ beyond the secondary barrier,  is composed  both by leakage $I_L [\frac{Sv}{h}]$ and
patient-scattered radiation $I_{ps}\,\, [\frac{Sv}{h}]$:

\be\label{1}
I_T=I_L+I_{ps}
\en

It is possible to measure $I_T$ (with phantom and with completely open collimator and completelly open multileafs) and $I_L$ (without phantom and with completely closed collimator and completelly closed multileafs) (\cite{NCRP151})

Then $I_{ps}$ will be given by:

\be\label{2}
I_{ps}=I_T-I_L
\en

Each measure  will be given by a reading ($L$ (leakage) and $L_T$ (total)) and the Natural Background ($L_{BG}$) as: 

\be\label{deft}
I_T= L_T-L_{BG}
\en

\be\label{defl}
I_L=L-L_{BG} \,.
\en

\subsubsection{Method 1 (Historical old method)}

According to this method only $I_T$ is measured, so that the total weekly dose-equivalent rate, call it $R'\, \,  [\frac{Sv}{week}]$, for a weekly workload $W \,\, [\frac{Gy}{week}]$, leakage-radiation  workload $W_L \,\, [\frac{Gy}{week}]$,  occupation factor $T$ and $\dot{D_o}\equiv$ absorbed-dose output rate at $1\, m$ in $ [\frac{Gy}{h}]$, is calculated by

\vspace{1cm}

a) Before IMRT- In this case $W=W_L$ i.e. there is an unique weekly workload, using (\ref{deft}), we have: 

\be\label{semIMRT}
R'= (L_{T}- L_{BG}) \frac{W}{\dot{D_o}} \, U\, T \, .
\en


\vspace{1cm}

b) With IMRT: $W$ is substituted by $W_L$ and we have:

\be\label{velhoW}
R'= (L_{T}- L_{BG}) \frac{W_L}{\dot{D_o}} \, U\, T\, .
\en

We remark that this equation comes from the equation used historically in which the weekly primary load $ W $ has been replaced by $ W_L$.

\subsubsection{Method 2 (new)}

Taking into account Eq. (\ref{1}),  the total weekly dose-equivalent rate, $R \, \,  [\frac{Sv}{week}]$, for a weekly workload $W \,\, [\frac{Gy}{week}]$, leakage-radiation  workload $W_L \,\, [\frac{Gy}{week}]$,  occupation factor $T$, use factor $U$ and $\dot{D_o}\equiv$ absorbed-dose output rate at $1\, m$ in $ [\frac{Gy}{h}]$, is given by (\cite{NCRP151})

\be\label{4}
R= \left\{I_{L}\frac{W_L}{\dot{D_o}}+ (I_T-I_L)\frac{W \,\, U}{\dot{D_o}}\right\}T\, ,
\en


Again, each measure  will be given by a reading ($L$ (leakage) and $L_T$ (total)) and the Natural Background ($L_{BG}$), so that, using (\ref{deft}) and (\ref{defl}),  the equation (\ref{4}) can be written as

\be\label{5}
R= \left\{(L- L_{BG}) \frac{W_L}{\dot{D_o}}+ (L_T-L)\frac{W \,\, U}{\dot{D_o}}\right\}T   \, .
\en

Note that 

\be\label{6}
\frac{W}{\dot{D_o}}=  \equiv t_{beam \, on}\,\,\, in \,[\frac{h}{week}]\,\,and \,\, we \,\, call\,\, it\,\,  t_{bo}
\en

and

\be\label{7}
\frac{W_L}{\dot{D_o}}=  \equiv t_{beam \, on - IMRT}\,\,\,in \,[\frac{h}{week}] \, \, and \,\, we \,\, call \,\, it  \,\, t_L\,\,;
\en

so we can write Eq. (\ref{5}) as

\be\label{novo}
R= \left\{(L- L_{BG}) \,\, t_L + (L_T-L) \,\,U \,\, t_{bo}\right\}T \, .
\en

The equation (\ref{novo}) is the heart of the new methodology proposed by the authors.

\subsubsection{Comparison between the two methods.}

Using definition (\ref{7}) we can write for the old method, Eq.(\ref{velhoW})

\be\label{8}
R'= (L_{T}- L_{BG}) \, U\, T \, t_L 
\en

and using  that: 

\be
L_T= L+ L_T -L
\en
 
we have

\be\label{velho}
R'= \left\{(L- L_{BG}) \,\, U\,\, t_L + (L_T - L) \,\,U \,\, t_L\right\}T
\en

In order to compare we write again the two equations (\ref{velho})(old method) and (\ref{novo})(new method)  together

\be\label{novoc}
R= \left\{(L- L_{BG}) \,\, t_L + (L_T-L) \,\,U \,\, t_{bo}\right\}T \,\,\,\,\,\,\,\,\,\, "new"
\en

\be\label{velhoc}
R'= \left\{(L- L_{BG}) \,\, U\,\, t_L + (L_T - L) \,\,U \,\, t_L\right\}T \,\,\,\,\,\,\,\,\,\, "old"
\en

The proposed equation is (\ref{novoc}) and it is deducted from the theory. We can notice in equation  (\ref{velhoc}) two problems: i) the scattered component (the second term of the RHS) is multiplied by $t_L$ instead of the correct value $t_{bo}$ (we know that the weekly scattered component is not altered by IMRT) and ii) the leakage component (the first term of the RHS)  is affected by the use factor $U$ even if it is  different from 1 (we know that the leakage radiation is always present for any Gantry orientation: U = 1). 

Then there will be differences in measuring dose rates according to one method or another. To analyze these differences we are going to consider two situations: the first is when the use factor of the secondary wall is the unit $ U =  1$ (call it  {\it "pure"}) and the second when $ U < 1 $ (call it  {\it "not pure"}).

\vspace{1cm}

{\bf A)Situation $U=1$.}

From Eqs. (\ref{novoc}) and (\ref{velhoc}), making $U=1$, we have

\be\label{excess}
R'-R = (L_T-L)\,\,(t_L - t_{bo})\,\, T
\en

Since on the Right Side of this equation each factor is positive, we have  $R '- R$ is always positive 

\be
R'-R = (L_T-L)\,\,(t_L - t_{bo})\,\, T > 0
\en

This  already is an indication that the old method, in this situation,  overestimates the doses.

Let's calculate  the relative excess $\frac{R'-R}{R}$. From Eqs. (\ref{excess}) and (\ref{novoc}) we obtain:

\be\label{excessr}
\frac{R'-R}{R}=\frac{t_L-t_{bo}}{\frac{L-L_{BG}}{L_T-L}\, t_L +t_{bo}}\,, \,\,\,\, valid \,\,  if \,\, \,L_T-L \neq 0\, .
\en

{\bf B)Situation $U<1$.}

From Eqs. (\ref{novoc}) and (\ref{velhoc}) we obtain

\be
R'-R= \left\{(L-L_{BG})\, (U-1)\, t_L \,+\, (L_T-L)\, U\, (t_L-t_{bo})\right\}\,\, T
\en

Investigating the sign of $R '-R$ found two cases:

\vspace{1cm}

{\bf Case B1)} "overestimated"

\be
R'-R \geq 0   \Longleftrightarrow 
\en

\be\label{b1}
\frac{L_T - L}{L-L_{BG}}\geq \frac{1-U}{U} \,.\,  \frac{t_L}{t_L-t_{bo}}\, ,  \,\,\,\, valid \,\,  if \,\, L-L_{BG} \neq 0 \,\,\, and \,\,\,t_L-t_{bo} \neq 0   
\en

or

\be\label{b12}
\frac{Scattered}{Leakage - BG}\geq \frac{1-U}{U} \,.\, \frac{t_L}{t_L-t_{bo}}\, , \,\,\,\, valid \,\,  if \,\, Leakage - BG \neq 0\,\,\, and \,\,\,t_L-t_{bo} \neq 0 \, .
\en

 {\bf Case B2)} "subestimated"

\be
R'-R < 0   \Longleftrightarrow 
\en

\be\label{b2}
\frac{L_T - L}{L-L_{BG}} < \frac{1-U}{U} \,.\, \frac{t_L}{t_L-t_{bo}}\, , \,\,\,\, valid \,\,  if \,\, L-L_{BG} \neq 0\,\,\, and \,\,\,t_L-t_{bo} \neq 0
\en

or

\be\label{b22}
\frac{Scattered}{Leakage- BG} < \frac{1-U}{U} \,.\, \frac{t_L}{t_L-t_{bo}}\, , \,\,\,\, valid \,\,  if \,\, Leakage - BG \neq 0\,\,\, and \,\,\,t_L-t_{bo} \neq 0  \, .
\en

Eq. (\ref{b1}) or (\ref{b12}) gives the conditions for $R'-R\geq 0$ i.e. $R'$ overestimate doses. Eq. (\ref{b2}) or (\ref{b22}) gives the conditions for $R'-R < 0$ i.e. $R'$ under-estimate doses.

\vspace{1cm}
For the relative difference we obtain: 
\be\label{26}
\frac{R'-R}{R}=\frac{(U-1) t_L +\frac{L_T-L}{L-L_{BG}} U (t_L - t_{bo})}{t_L+\frac{L_T-L}{L-L_{BG}}U t_{bo}}\, , \, \,\,\,\, valid \,\,  if \,\, L-L_{BG} \neq 0\, . 
\en

\section{Results}

In this section we particularize the situations and cases found in the last section, for concrete real examples with real values of parameters and verify them experimentally.

\vspace{1cm}

{\bf A)Situation $U=1$.} Examples

\vspace{1cm}

\underline{\bf BUNKER 1)}  \, For a LINAC of 10 MV with primary workload $W=1100\frac{Gy}{week}$ and secondary IMRT leakage load $W_L=3300\frac{Gy}{week}$ and in a nominal absorbed-dose output rate $\dot{D_{o}}=360\frac{Gy}{h}$ we have:

\vspace{1cm}

\be
t_{bo}=\frac{1100\frac{Gy}{week}}{360\frac{Gy}{h}}= \frac{10}{3}\frac{h}{week} \cong\ 3,06 \frac{h}{week}
\en

\be
t_{L}=\frac{3300\frac{Gy}{week}}{360\frac{Gy}{h}}\frac{22}{3}\frac{h}{week}    \cong 9,17 \frac{h}{week}
\en

\begin{figure}[t!]
\begin{center}
\includegraphics[width=12cm]{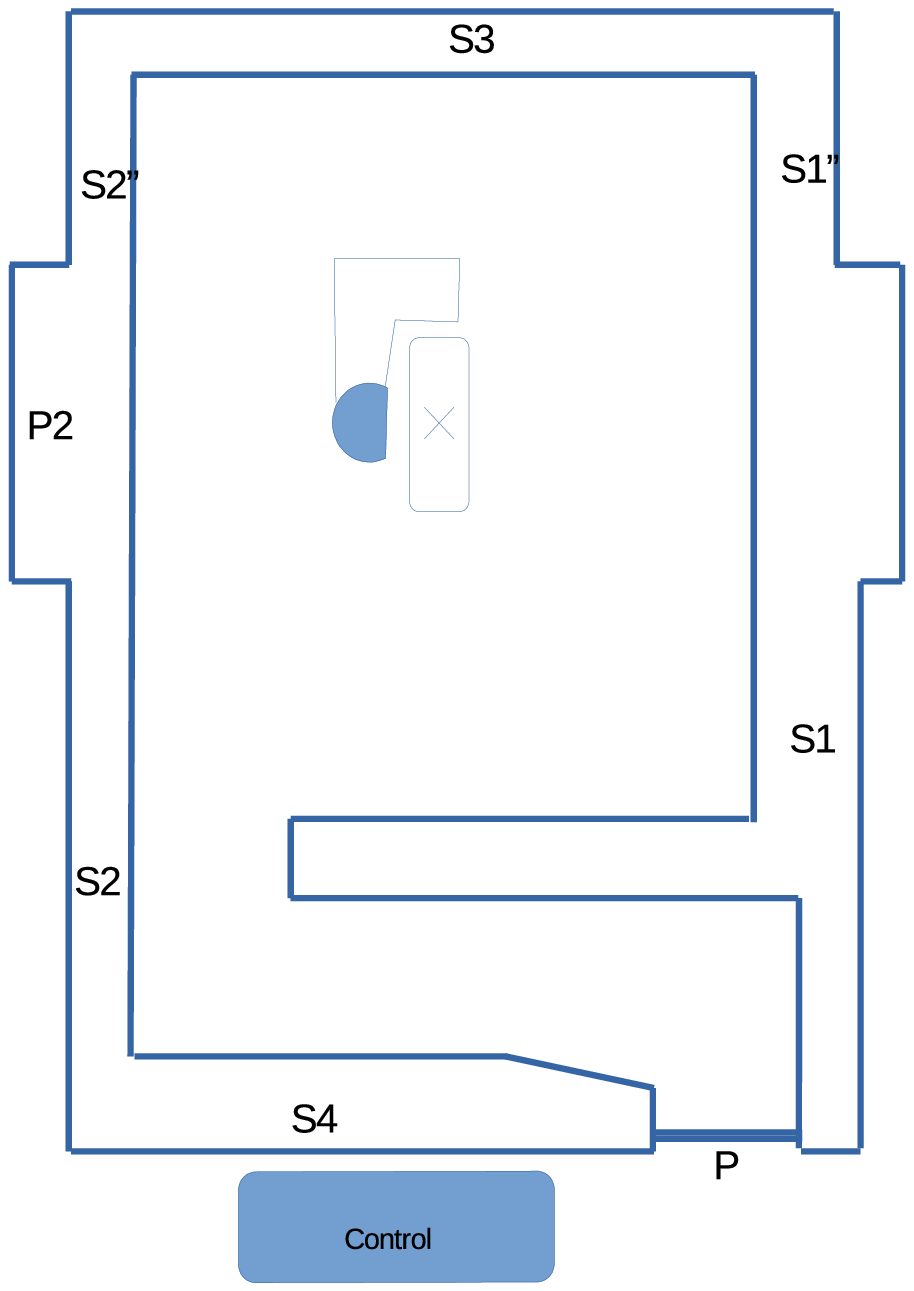} 
\caption{Bunker 1. The neighborhoods of each wall are indicated in the table I}
\label{bunker21}
\end{center}
\end{figure}

\begin{table}[t!]
\begin{tabular}{||c|c|c|c|c|c|c|c|c|c|c||}\hline\hline

\ Wall & Destination & T & U & Area &$L_{T} [\frac{\mu Sv}{h}]$ & $L [\frac{\mu Sv}{h}]$& $L_{BG}[\frac{\mu Sv}{h}]$ &$ R' \frac{\mu Sv}{week} $ & $R\frac{\mu Sv}{week}$ & $ Difference = \frac{R'-R}{R}\times 100$
\\ \hline

S1   & Treatment Room           &$ \frac{1}{2}$ &$1$& Uncontrolled &$5.01$ & $0.19 $&$0.09$ & $\color{red}22.55$&$ 7.82$             & 188 \%\\ \hline

S1'   & Treatment Room          & $\frac{1}{2}$ &$1$& Uncontrolled &$15.1$ & $0.21$ &$0.09$ & $\color{red}68.80$& $\color{red}23.30$ & 190 \%\\ \hline

S2   & Rest  area               & $1$           &$1$& Uncontrolled &$5.49$ & $0.24$ &$0.09$ & $\color{red}49.50$& $ 17.42$           & 184 \% \\ \hline

S2'   & Rest area               & $1$           &$1$& Uncontrolled &$2.30$ & $0.12 $&$0.09$ & $20.26$           &$6.94$              & 192 \%\\ \hline

P   & Door                      & $\frac{1}{8} $& $1$& Uncontrolled&$8.0$  & $NM$ &$0.09$ & $9.064 $& $-$& - \%\\ \hline



Ts  & Reception                 &$1 $           &$1$ &Uncontrolled&$0.31$   &$0.21$  & $0.09$&$2.017     $       & $1.406 $            &43 \%\\ \hline



\hline\hline 
\end{tabular}
\caption{Bunker 1. Measurements for secondaries walls with Ionization pressurized chamber Ludlum 9DP serial number  25009346. This bunker houses a 10 MV LINAC with primary workload $W=1100\frac{Gy}{week}$ and secondary IMRT leakage load $W_L=3300\frac{Gy}{week}$ and in a nominal absorbed-dose output rate $\dot{D_{o}}=360\frac{Gy}{h}$. The Old method says that 3 walls are not safe (S1, S1' and S2). But the new method says that only S1' is not good from the point of view of radiological protection. Several walls (S2', P and Ts) were safe according to the old method, which as we know, over-sizes in this situation ($U=1$), then it is not necessary to apply the new method that is, the leak component $L$ need not be measured (NM). We did this in the case of wall P . In other words, for the situation $U = 1$, we can leave the new method only to be applied to the hot points of the old method.}
\end{table}\label{tabla2}

Measures with Ionization pressurized chamber Ludlum 9DP serial number  25009346, we obtained for the following walls:

\vspace{1cm}

-{\bf S1}: Treatment room. Occupancy factor $T=\frac{1}{2}$. Uncontrolled area with contribution from another source. Then limit: $P=10\frac{\mu Sv}{week}$.

\vspace{1cm}

$L_T=5,01 \frac{\mu Sv}{h}$

$L=0,19\frac{\mu Sv}{h}$

$L_{BG}=0,09 \frac{\mu Sv}{h}$

\vspace{0.5cm}

Then from Eqs. (\ref{excess}) and (\ref{excessr}) we have

\be
R'-R= (5,01\frac{\mu Sv}{h} - 0,19\frac{\mu Sv}{h})(9,17\frac{h}{week}-3,06\frac{h}{week}) \frac{1}{2}=   14,725 \frac{\mu Sv}{week}
\en

\be
\frac{R'-R}{R}=\frac{t_L-t_{bo}}{\frac{L-L_{BG}}{L_T-L}\, t_L +t_{bo}}\,= \frac{9,17\frac{h}{week}-3,06\frac{h}{week}}{\frac{0,19\frac{\mu Sv}{h}-0,09\frac{\mu Sv}{h}}{5,01\frac{\mu Sv}{h} - 0,19\frac{\mu Sv}{h}}\, 9,17\frac{h}{week} +3,06\frac{h}{week}}\cong 1,88
\en
It means

\vspace{1cm}

\be
R'\cong 2,88 R
\en 
 
Or equivalently 

\vspace{1cm}

$R'= 22,55 \frac{\mu Sv}{week}$ 

$R = 7,822\frac{\mu Sv}{week}$ \, , 

\vspace{1cm}

and then  the old method is giving  $\cong$ 188 $\% $ excess for this wall.

In this example, very interesting,  the old method indicates a dose rate that exceeds in $\cong 125\% $ the allowed limit for an uncontrolled area where contributes two sources (i.e $P=10\frac{\mu Sv}{week}$), while the new method indicates a lower dose rate than this allowed limit. In other words, the old method is incorrectly condemning  this wall. Then the old method must be discarded for this wall.

\vspace{1cm}

-{\bf S2}: Rest area. Uncontrolled area with $T=1$.

\vspace{1cm}

$L_T=5,49 \frac{\mu Sv}{h}$

$L=0,24\frac{\mu Sv}{h}$

$L_{BG}=0,09 \frac{\mu Sv}{h}$

\vspace{0.5cm}

Then from Eqs. (\ref{excess}) and (\ref{excessr}) we have

\be
R'-R= (5,49\frac{\mu Sv}{h} - 0,24\frac{\mu Sv}{h})(9,17\frac{h}{week}-3,06\frac{h}{week}) =   32,078 \frac{\mu Sv}{week}
\en

\be
\frac{R'-R}{R}=\frac{t_L-t_{bo}}{\frac{L-L_{BG}}{L_T-L}\, t_L +t_{bo}}\,= \frac{9,17\frac{h}{week}-3,06\frac{h}{week}}{\frac{0,24\frac{\mu Sv}{h}-0,09\frac{\mu Sv}{h}}{5,49\frac{\mu Sv}{h} - 0,24\frac{\mu Sv}{h}}\, 9,17\frac{h}{week} +3,06\frac{h}{week}}\cong 1,84
\en
It means

\vspace{1cm}

\be
R'\cong 2,84 R
\en 
 
Or equivalently 

\vspace{1cm}

$R'= 49,50 \frac{\mu Sv}{week}$ 

$R = 17,42\frac{\mu Sv}{week}$ \, , 

\vspace{1cm}

and then  the old method is giving  $\cong$ 184 $\% $ excess for this wall.

We see that  the old method indicates a dose rate that exceeds in $\cong 148\% $ the allowed limit for an uncontrolled area ($P=20\frac{\mu Sv}{week}$), while the new method indicates a lower dose rate than this allowed limit. In other words, the old method again is incorrectly condemning  a wall. Then the old method must be discarded for this wall.

\vspace{1cm}

-{\bf S1'}: Treatment room. Occupancy factor $T=\frac{1}{2}$. Uncontrolled area with contribution from another source. Then limit: $P=10\frac{\mu Sv}{week}$.

\vspace{1cm}

$L_T=15,10 \frac{\mu Sv}{h}$

$L=0,21\frac{\mu Sv}{h}$

$L_{BG}=0,09 \frac{\mu Sv}{h}$

\vspace{0.5cm}

Then from Eqs. (\ref{excess}) and (\ref{excessr}) we have

\be
R'-R= (15,10\frac{\mu Sv}{h} - 0,21\frac{\mu Sv}{h})(9,17\frac{h}{week}-3,06\frac{h}{week}) \frac{1}{2}=   32,078 \frac{\mu Sv}{week}
\en

\be
\frac{R'-R}{R}=\frac{t_L-t_{bo}}{\frac{L-L_{BG}}{L_T-L}\, t_L +t_{bo}}\,= \frac{9,17\frac{h}{week}-3,06\frac{h}{week}}{\frac{0,21\frac{\mu Sv}{h}-0,09\frac{\mu Sv}{h}}{15,10\frac{\mu Sv}{h} - 0,21\frac{\mu Sv}{h}}\, 9,17\frac{h}{week} +3,06\frac{h}{week}}\cong 1,95
\en
It means

\vspace{1cm}

\be
R'\cong 2,95 R
\en 
 
Or equivalently 

\vspace{1cm}

$R'= 68,82 \frac{\mu Sv}{week}$ 

$R = 23,33\frac{\mu Sv}{week}$ \, , 

\vspace{1cm}

and then  the old method is giving  $\cong$ 195 $\% $ excess for this wall.

For this wall the old method indicates a dose rate that exceeds in $\cong 588\% $ the allowed limit for an uncontrolled area where contributes two sources (i.e $P=10\frac{\mu Sv}{week}$), while the new method indicates  a dose rate exceeding $\cong 133\% $ this allowed limit. In other words, both methods are condemning  this wall and it will have to be reformulated. Even so, the new method requires increasing the thickness of the wall by only approximately 1.2 HVL when the old method asks approximately 2.8 HVL.

\vspace{1cm}

-{\bf S2'}: Rest area. Uncontrolled area with $T=1$. Allowed Limit $P=20\frac{\mu Sv}{week}$

\vspace{1cm}

$L_T=2,30 \frac{\mu Sv}{h}$

$L=0,12\frac{\mu Sv}{h}$

$L_{BG}=0,09 \frac{\mu Sv}{h}$

\vspace{0.5cm}

Then from Eqs. (\ref{excess}) and (\ref{excessr}) we have

\be
R'-R= (2,30\frac{\mu Sv}{h} - 0,12\frac{\mu Sv}{h})(9,17\frac{h}{week}-3,06\frac{h}{week}) =   13,32 \frac{\mu Sv}{week}
\en

\be
\frac{R'-R}{R}=\frac{t_L-t_{bo}}{\frac{L-L_{BG}}{L_T-L}\, t_L +t_{bo}}\,= \frac{9,17\frac{h}{week}-3,06\frac{h}{week}}{\frac{0,12\frac{\mu Sv}{h}-0,09\frac{\mu Sv}{h}}{2,30\frac{\mu Sv}{h} - 0,12\frac{\mu Sv}{h}}\, 9,17\frac{h}{week} +3,06\frac{h}{week}}\cong 1,92
\en
It means

\vspace{1cm}

\be
R'\cong 2,92 R
\en 
 
Or equivalently 

\vspace{1cm}

$R'= 20,26 \frac{\mu Sv}{week}$ 

$R = 6,94\frac{\mu Sv}{week}$ \, , 

\vspace{1cm}

and then  the old method is giving  $\cong$ 192 $\% $ excess for this wall.

We see that  the old method indicates a dose rate that exceeds in only $\cong 1,3\% $ the allowed limit for an uncontrolled area ($P=20\frac{\mu Sv}{week}$) which is well within the margin of error (we accept up to 20 $\% $ ), while the new method indicates a lower dose rate than this allowed limit. In other words, both methods indicates an acceptable dose rate for this wall. 
The results for all secondary walls of this bunker (Fig. \ref{bunker21} ) are summarized in Table I.

\vspace{1cm}

\underline{\bf BUNKER 2)}  \, For a LINAC of 10 MV with primary workload $W=1200\frac{Gy}{week}$ and secondary IMRT leakage load $W_L=2640\frac{Gy}{week}$ and in a nominal absorbed-dose output rate $\dot{D_{o}}=360\frac{Gy}{h}$ we have:

\vspace{1cm}

\be
t_{bo}=\frac{1200\frac{Gy}{week}}{360\frac{Gy}{h}}= \frac{10}{3}\frac{h}{week} \cong\ 3,33 \frac{h}{week}
\en

\be
t_{L}=\frac{2640\frac{Gy}{week}}{360\frac{Gy}{h}}\frac{22}{3}\frac{h}{week}    \cong 7,33 \frac{h}{week}
\en

\begin{figure}[t!]
\begin{center}
\includegraphics[width=12cm]{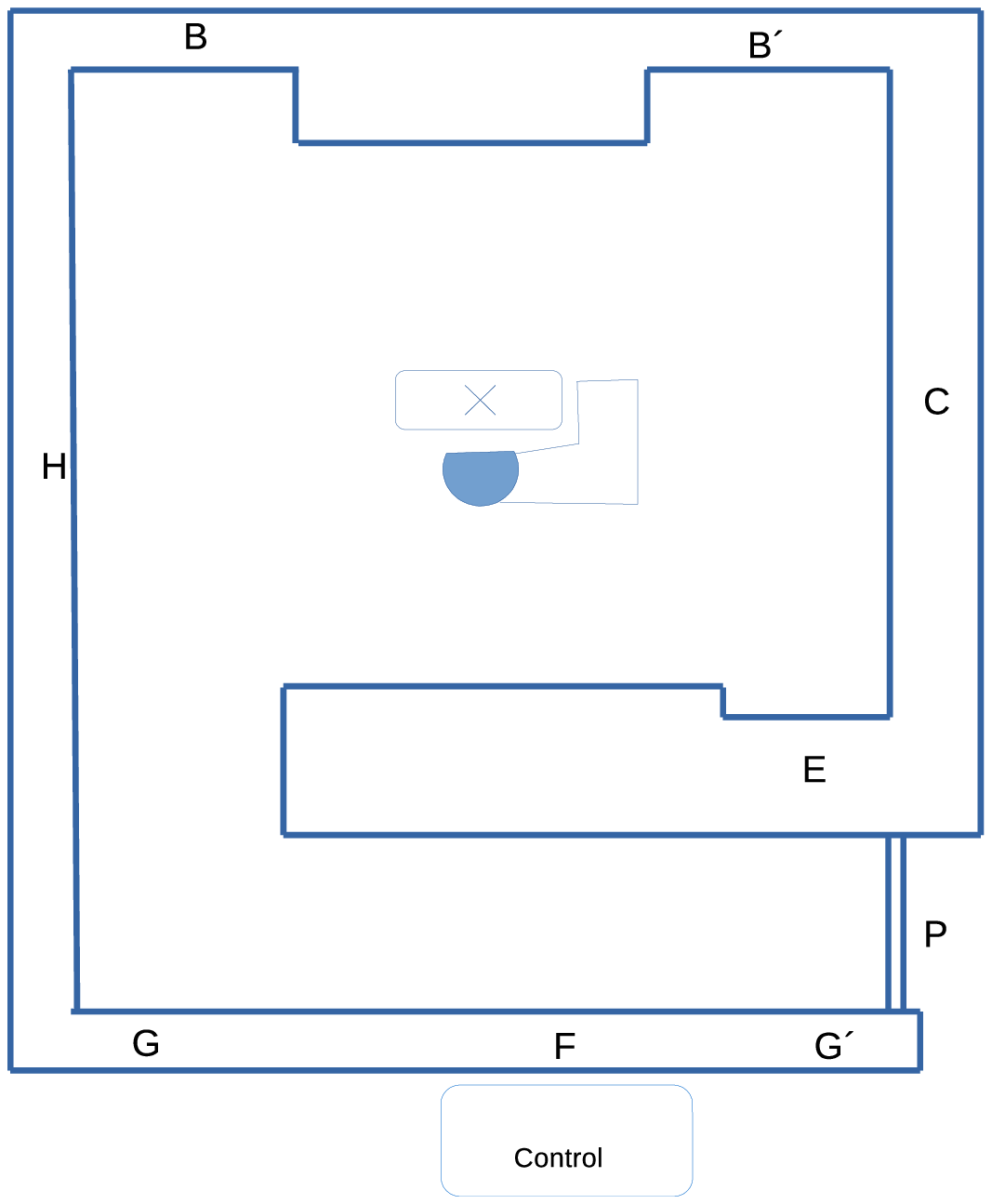} 
\caption{Bunker 2. The neighborhoods of each wall are indicated in the table II.}
\label{bunker1}
\end{center}
\end{figure}

\begin{figure}[t!]
\begin{center}
\includegraphics[width=12cm]{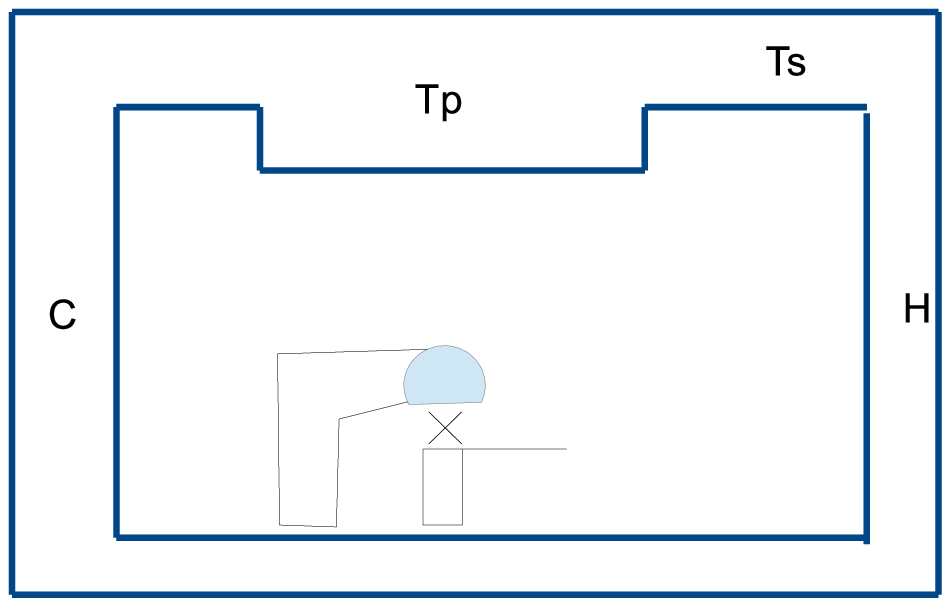} 
\caption{Bunker 2 Cut. The neighborhoods of each wall are indicated in the table II.}
\label{bunker1T}
\end{center}
\end{figure}

\begin{table}[t!]
\begin{tabular}{||c|c|c|c|c|c|c|c|c|c|c||}\hline\hline

\ Wall & Destination & T & U & Area &$L_{T} [\frac{\mu Sv}{h}]$ & $L [\frac{\mu Sv}{h}]$& $L_{BG}[\frac{\mu Sv}{h}]$ &$ R' \frac{\mu Sv}{week} $ & $R\frac{\mu Sv}{week}$ & $ Difference = \frac{R'-R}{R}\times 100$
\\ \hline

Ts   & Parking          &$ \frac{1}{8}$ &$1$& Uncontrolled &$36.90 $ & $0.80 $&$0.32$ & $\color{red}33.53$ &$ 15.48$ & 117 \%\\ \hline

B   & Subsoil           &  $\frac{1}{8}$ &$1$ & Uncontrolled   &$0.40$ & $0.40$  &$0.32$ & $0.07$ & $0.07$ & 0 \%\\ \hline

C   & Treatment Room    & $\frac{1}{4}$ & $1$  & Uncontrolled &$4.00 $ & $3.40$ & $0.32$ & $6.74$ & $ 6.14$& 9.8 \% \\ \hline

G   & Technical area    & $\frac{1}{8} $ &$1$ & Uncontrolled &$8.60$  & $0.41 $&$0.32$ & $7.59$ &$3.49$  & 117 \%\\ \hline

H   & Sidewalk          & $\frac{1}{8} $ & $1$ & Uncontrolled&$0.40$  & $0.40$ &$0.32$ & $0.07 $& $0.07$& 0 \%\\ \hline

G'   & Circulation      & $\frac{1}{8} $ &$1$ & Uncontrolled  &$3.60$  & $0.50$ &$0.32$ & $3.01 $&$1.46$ & 106 \%\\ \hline





\hline\hline 
\end{tabular}
\label{tabla1}
\caption{Bunker 2. Measurements for secondaries walls with Ionization pressurized chamber Ludlum 9DP serial number  25018180. This bunker houses a LINAC of 10 MV with primary workload $W=1200\frac{Gy}{week}$ and secondary IMRT leakage load $W_L=2640\frac{Gy}{week}$ and in a nominal absorbed-dose output rate $\dot{D_{o}}=360\frac{Gy}{h}$.  The old method is incorrectly condemning  the wall  Ts (Roof) while the new method indicates a  dose rate lower than the allowed limit. For the others walls both methods give adequate values.}
\end{table}

Measures with Ionization pressurized chamber Ludlum 9DP serial number  25018180, we obtained for

\vspace{1cm}

-{\bf Ts Roof}: Parking and access ramp. Uncontrolled area with $ T=\frac{1}{8} $.

\vspace{1cm}

$L_T=36,9 \frac{\mu Sv}{h}$

$L=0,8\frac{\mu Sv}{h}$

$L_{BG}=0,322 \frac{\mu Sv}{h}$

Then from Eqs. (\ref{excess}) and (\ref{excessr}) we have

\be
R'-R= (36,9\frac{\mu Sv}{h} - 0,8\frac{\mu Sv}{h})(\frac{22}{3}\frac{h}{week}-\frac{10}{3}\frac{h}{week}) \frac{1}{8}=   18,05 \frac{\mu Sv}{week}
\en

\be
\frac{R'-R}{R}=\frac{t_L-t_{bo}}{\frac{L-L_{BG}}{L_T-L}\, t_L +t_{bo}}\,= \frac{\frac{22}{3}\frac{h}{week}-\frac{10}{3}\frac{h}{week}}{\frac{0,8\frac{\mu Sv}{h}-0,322\frac{\mu Sv}{h}}{36,9\frac{\mu Sv}{h} - 0,8\frac{\mu Sv}{h}}\, \frac{22}{3}\frac{h}{week} +\frac{10}{3}\frac{h}{week}}\cong 1,166
\en
It means

\vspace{1cm}

\be
R'\cong 2,166 R
\en 
 
Or equivalently 

\vspace{1cm}

$R'= 33,53 \frac{\mu Sv}{week}$ 

$R = 15,48\frac{\mu Sv}{week}$ \, , 

\vspace{1cm}

and then  the old method is giving  $\cong$ 117 $\% $ excess for this wall.

In this example, very interesting,  the old method indicates a dose rate that exceeds in $\cong 68\% $ the allowed limit for uncontrolled area ($20\frac{\mu Sv}{week}$) wall {\bf Ts Roof} , while the new method indicates a lower dose rate than the allowed limit. In other words, the old method is incorrectly condemning  the wall {\bf Ts Roof}. Then the old method must be discarded for this wall. For the others walls (Figs. \ref{bunker1} and \ref{bunker1T} ) both methods give adequate values, see Table II .

\vspace{1cm}

\underline{\bf BUNKER 3)}  \, For a LINAC of 15 MV with primary workload $W=1150\frac{Gy}{week}$ and secondary IMRT leakage load $W_L=1750\frac{Gy}{week}$ and in a nominal absorbed-dose output rate $\dot{D_{o}}=360\frac{Gy}{h}$ we have:

\vspace{1cm}

\be
t_{bo}=\frac{1150\frac{Gy}{week}}{360\frac{Gy}{h}}\cong\ 3,19 \frac{h}{week}
\en

\be
t_{L}=\frac{1750\frac{Gy}{week}}{360\frac{Gy}{h}}    \cong 4,86 \frac{h}{week}
\en

Measures with Ionization pressurized chamber Ludlum 9DP serial number  25018180, we obtained for

-{\bf M2}: Chemiotherapy room. Occupancy factor $T=1$. Uncontrolled area, limit: $P=20\frac{\mu Sv}{week}$.

\vspace{1cm}

$L_T=6,0 \frac{\mu Sv}{h}$

$L=0,8\frac{\mu Sv}{h}$

$L_{BG}=0,53 \frac{\mu Sv}{h}$

\vspace{0.5cm}

Then from Eqs. (\ref{excess}) and (\ref{excessr}) we have

\be
R'-R= (6,0\frac{\mu Sv}{h} - 0,8\frac{\mu Sv}{h})(4,86\frac{h}{week}-3,19\frac{h}{week}) =   8,68 \frac{\mu Sv}{week}
\en

\be
\frac{R'-R}{R}=\frac{t_L-t_{bo}}{\frac{L-L_{BG}}{L_T-L}\, t_L +t_{bo}}\,= \frac{4,86\frac{h}{week}-3,19\frac{h}{week}}{\frac{0,8\frac{\mu Sv}{h}-0,53\frac{\mu Sv}{h}}{6,0\frac{\mu Sv}{h} - 0,8\frac{\mu Sv}{h}}\, 4,86\frac{h}{week} + 3,19\frac{h}{week}}\cong 0,485
\en
It means

\vspace{1cm}

\be
R'\cong 1,485 R
\en 
 
Or equivalently 

\vspace{1cm}

$R'= 26,58 \frac{\mu Sv}{week}$ 

$R = 17,9\frac{\mu Sv}{week}$ \, , 

\vspace{1cm}

and then  the old method is giving  $\cong$ 48 $\% $ excess for this wall.

In this example, very interesting,  the old method indicates a dose rate that exceeds in $\cong 33\% $ the allowed limit for an uncontrolled area (i.e $P=20\frac{\mu Sv}{week}$), while the new method indicates a lower dose rate than this allowed limit. In other words, the old method is incorrectly condemning  this wall. Then the old method must be discarded for this wall.

\begin{figure}[t!]
\begin{center}
\includegraphics[width=12cm]{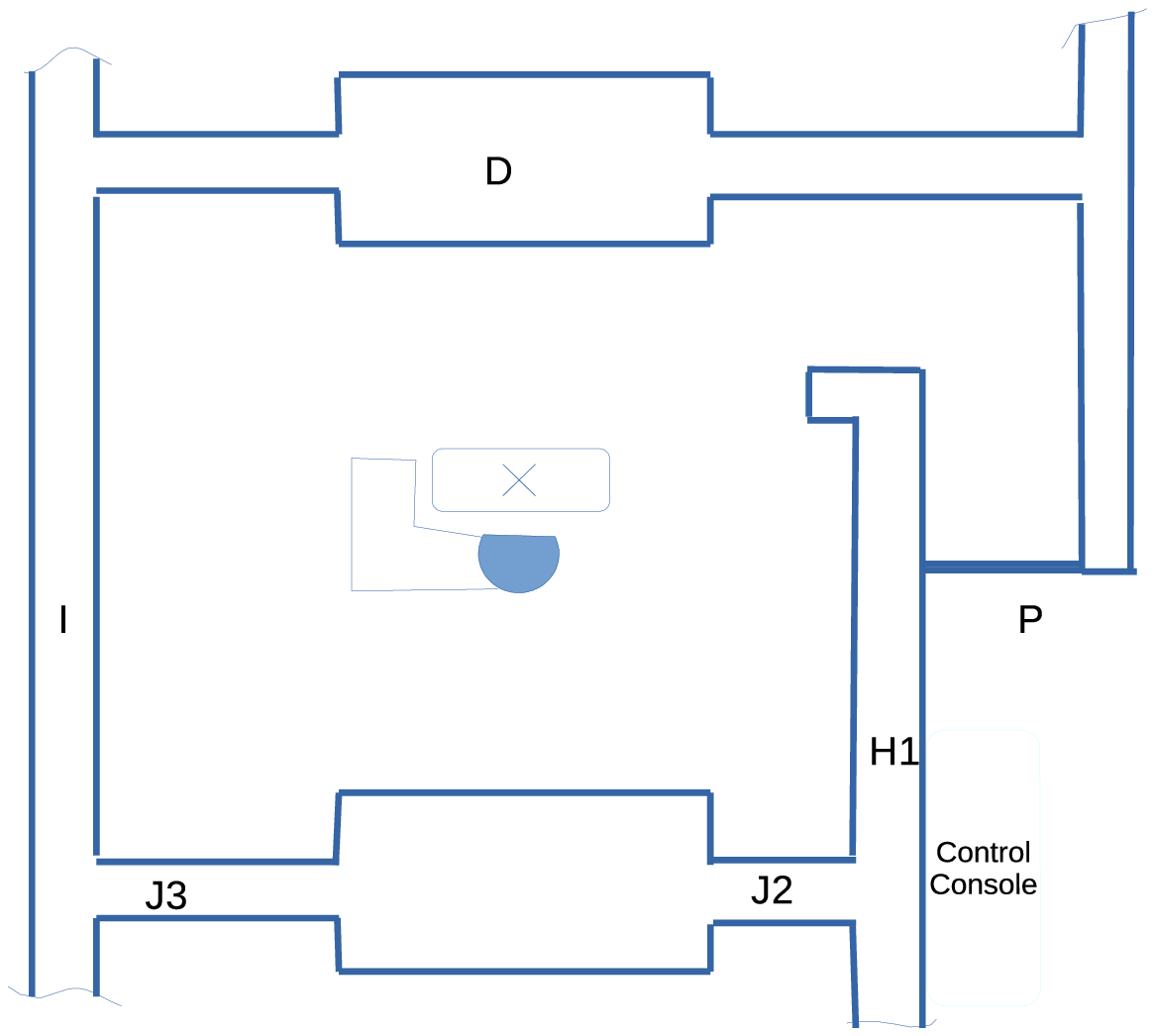} 
\caption{Bunker 3. The neighborhoods of each wall are indicated in the table III.}
\label{bunker3}
\end{center}
\end{figure}

\begin{figure}[t!]
\begin{center}
\includegraphics[width=12cm]{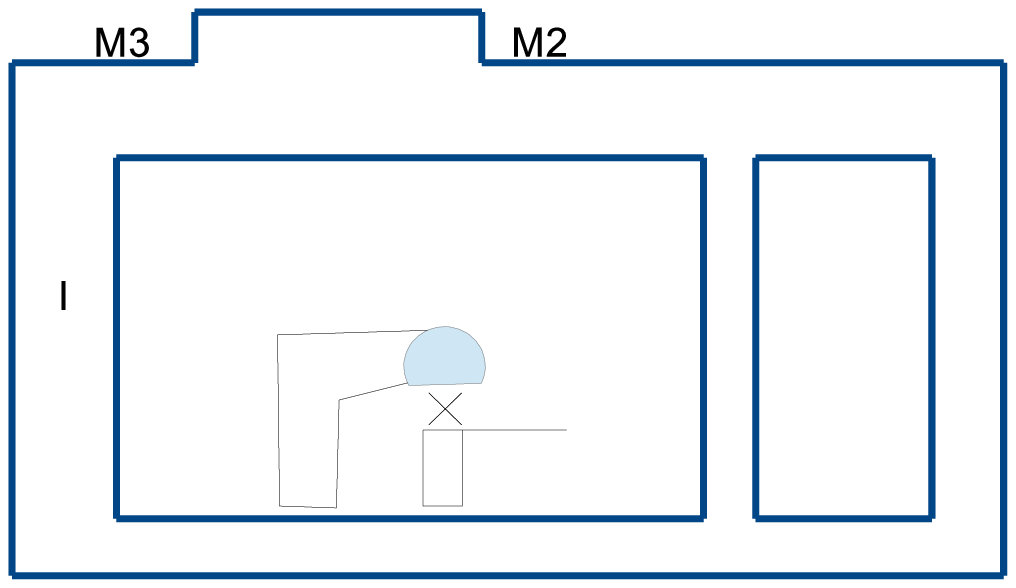} 
\caption{Bunker 3 Cut. The neighborhoods of each wall are indicated in the table III.}
\label{bunker3Tcut}
\end{center}
\end{figure}

For all the others secondary walls of this bunker (Fig. \ref{bunker3} and \ref{bunker3Tcut} ) the old method give adequate values, then it was no necessary apply the new method because in this case, we know, the old method overestimates.  In other words, for the cases $U=1 $, we are using the new method to be applied to the hot points of the old method. See Table III.

\begin{table}[t!]
\begin{tabular}{||c|c|c|c|c|c|c|c|c|c|c||}\hline\hline

\ Wall & Destination & T & U & Area &$L_{T} [\frac{\mu Sv}{h}]$ & $L [\frac{\mu Sv}{h}]$& $L_{BG}[\frac{\mu Sv}{h}]$ &$ R' \frac{\mu Sv}{week} $ & $R\frac{\mu Sv}{week}$ & $ Difference = \frac{R'-R}{R}\times 100$
\\ \hline

H1   & Console                 &$1$            &$1$& Controlled  &$8.2$   & $NM $&$0.53$ & $37.29$& $ - $& $- \%$\\ \hline

J2   & Treatment Room         & $\frac{1}{2}$ &$1$& Uncontrolled &$8.6$ & $NM$ &$0.53$ & $19.62$ & $ -$ & $- \%$\\ \hline

J3   & Treatment Room         & $\frac{1}{2}$ &$1$& Uncontrolled &$4.2$ & $NM$ &$0.53$ & $8.92$& $ - $  & $- \%$ \\ \hline

I   &  Circulation            & $\frac{1}{16}$&$1$& Uncontrolled &$24.7$ & $NM $&$0.53$ & $7.34$ &  $- $  & $- \%$ \\ \hline

P   & Door                    & $\frac{1}{8} $&$1$&Controlled&    $16.7$ & $NM$ &$0.53$ & $13.23 $& $-$ & $- \%$\\ \hline

M2  & Chemiotherapy          &   $1$         & $1$& Uncontrolled  &$6.0$ & $0.8$&$0.53$ & $\color{red}26.59$&$17.93$ & $48,3 \%$\\ \hline

M3   & Technical Area        & $\frac{1}{16}$&$1$ & Uncontrolled&$32.4$ & $NM$ & $0.53$ &$9.68$ &  $-$ & $- \%$\\ \hline




\hline\hline 
\end{tabular}
\label{tabla3}
\caption{Bunker 3. Measurements for secondaries walls with Ionization pressurized chamber Ludlum 9DP serial number  25018180. This bunker houses a LINAC of 15 MV with primary workload $W=1150\frac{Gy}{week}$ and secondary IMRT leakage load $W_L=1750\frac{Gy}{week}$ and in a nominal absorbed-dose output rate $\dot{D_{o}}=360\frac{Gy}{h}$. The old method says that the  wall M2 is not safe because the dose rate ($26.59 \frac{\mu Sv}{week}$) exceeds limit ($20 \frac{\mu Sv}{week}$). However, the new method says that is not the case: the correct dose rate is $17,93 \frac{\mu Sv}{week}$ which is less than the limit. The other secondary walls were correct according to the old method, which, as we know, super-dimensionates in this case, then it was not necessary to apply the new method that is, the leak component $L$ was not measured (NM). In other words, for the cases $U=1 $, we are using the new method to be applied to the hot points of the old method.}
\end{table}

\vspace{1cm}

\underline{\bf BUNKER 4)}  \, For a LINAC of 10 MV with primary workload $W=1000\frac{Gy}{week}$ and secondary IMRT leakage load $W_L=2200\frac{Gy}{week}$ and in a nominal absorbed-dose output rate $\dot{D_{o}}=360\frac{Gy}{h}$ we have:

\vspace{1cm}

\be
t_{bo}=\frac{1000\frac{Gy}{week}}{360\frac{Gy}{h}}\cong\ 2,78 \frac{h}{week}
\en

\be
t_{L}=\frac{2200\frac{Gy}{week}}{360\frac{Gy}{h}}    \cong 6,11 \frac{h}{week}
\en

Measures with Ionization pressurized chamber Ludlum 9DP serial number  25009318 , we obtained for:

-{\bf M1}: Console other LINAC. Occupancy factor $T=\frac{1}{2}$. Uncontrolled area with contribution from another source. Then limit: $P=10\frac{\mu Sv}{week}$.

\vspace{1cm}

$L_T=2,70 \frac{\mu Sv}{h}$ 

$L=0,09\frac{\mu Sv}{h}$

$L_{BG}=0,01 \frac{\mu Sv}{h}$

\vspace{0.5cm}

Then from Eqs. (\ref{excess}) and (\ref{excessr}) we have

\be
R'-R= (2,7\frac{\mu Sv}{h} - 0,09\frac{\mu Sv}{h})(6,11\frac{h}{week}-2,78\frac{h}{week}) =   8,69 \frac{\mu Sv}{week}
\en

\be
\frac{R'-R}{R}=\frac{t_L-t_{bo}}{\frac{L-L_{BG}}{L_T-L}\, t_L +t_{bo}}\,= \frac{6,11\frac{h}{week}-2,78\frac{h}{week}}{\frac{0,09\frac{\mu Sv}{h}-0,01\frac{\mu Sv}{h}}{2,70\frac{\mu Sv}{h} - 0,09\frac{\mu Sv}{h}}\, 6,11\frac{h}{week} + 2,78\frac{h}{week}}\cong 1,12
\en
It means

\vspace{1cm}

\be
R'\cong 2,12 R
\en 
 
Or equivalently 

\vspace{1cm}

$R'= 16,44 \frac{\mu Sv}{week}$ 

$R = 7,74\frac{\mu Sv}{week}$ \, , 

\vspace{1cm}

and then  the old method is giving  $\cong$ 112 $\% $ excess for this wall.

In this example, very interesting,  the old method indicates a dose rate that exceeds in $\cong 64\% $ the allowed limit for an uncontrolled area with contribution of another source (i.e $P=\frac{20\frac{\mu Sv}{week}}{2}=10\frac{\mu Sv}{week}$), while the new method indicates a lower dose rate than this allowed limit. In other words, the old method is incorrectly condemning  this wall. Then the old method must be discarded for this wall.

\begin{figure}[h!]
\begin{center}
\includegraphics[width=12cm]{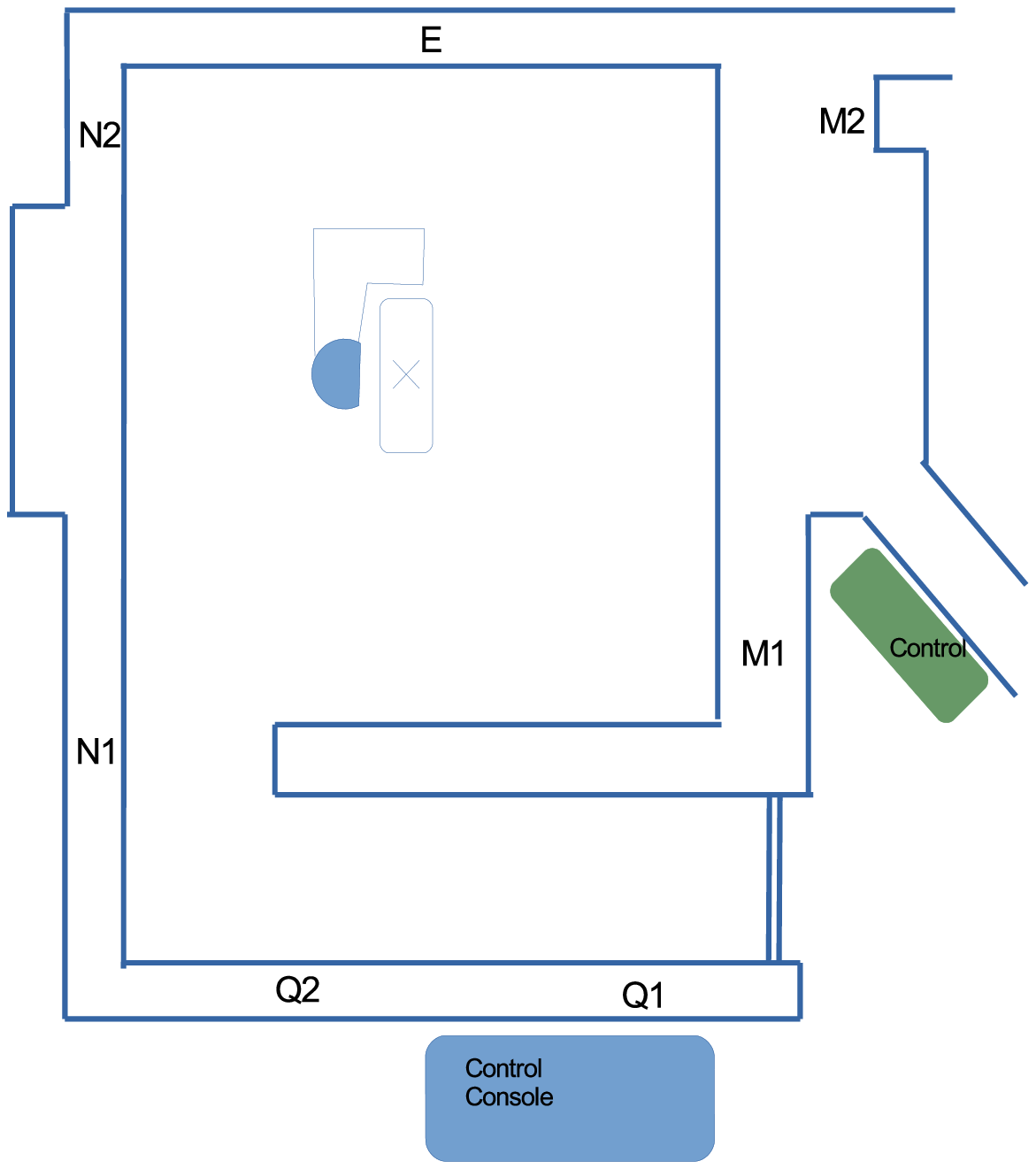}
\caption{Bunker 4. The neighborhoods of each wall are indicated in the table IV.}
\label{bunker4}
\end{center}
\end{figure}

For all the others secondary walls (Fig. \ref{bunker4} )the old method give adequate values, then it was no necessary apply the new method because in this case, we know, the old method overestimates.  In other words, for the cases $U=1 $, we are using the new method to be applied to the hot points of the old method. See Table IV.

\begin{table}[t!]
\begin{tabular}{||c|c|c|c|c|c|c|c|c|c|c||}\hline\hline

\ Wall & Destination & T & U & Area &$L_{T} [\frac{\mu Sv}{h}]$ & $L [\frac{\mu Sv}{h}]$& $L_{BG}[\frac{\mu Sv}{h}]$ &$ R' \frac{\mu Sv}{week} $ & $R\frac{\mu Sv}{week}$ & $ Difference = \frac{R'-R}{R}\times 100$
\\ \hline

M1   & Console neighbor LINAC &$1$            &$1$& Uncontrolled  &$2.7$   & $0.09 $&$0.01$ & $\color{red}16.44$& $7.74$& $112 \%$\\ \hline

D   & Door                    & $\frac{1}{2}$ &$1$& Uncontrolled &$0.45$ & $NM$ &$0.01$ & $1.34$ & $ -$ & $- \%$\\ \hline

M2   & Treatment Room         & $1$    &$1$      & Uncontrolled &$0.1$ & $NM$ &$0.01$ & $0.55$& $ - $  & $- \%$ \\ \hline

N1   &  Power station         & $\frac{1}{5}$&$1$& Uncontrolled &$4.7$ & $NM $&$0.01$ & $5.73$ &  $- $  & $- \%$ \\ \hline

N2   & Courtyard/P.station    & $\frac{1}{5} $&$1$&Uncontrolled&    $3.84$ & $NM$ &$0.01$ & $4.68$& $-$ & $- \%$\\ \hline

P  & Courtyard                &$\frac{1}{6}$ & $1$& Uncontrolled  &$0.15$ & $NM$&$0.01$ & $0.14$& $-$ & $- \%$\\ \hline

Q1   & Console                & $1$         & $1$ & Uncontrolled&$0.01 (BG)$ & $NM$ & $0.01$ &$0$ &  $-$ & $- \%$\\ \hline




\hline\hline 
\end{tabular}
\label{tabla4}
\caption{Bunker 4. Measurements for secondaries walls with Ionization pressurized chamber Ludlum 9DP serial number  25009318. This bunker houses a LINAC of 10 MV with primary workload $W=1000\frac{Gy}{week}$ and secondary IMRT leakage load $W_L=2200\frac{Gy}{week}$ and in a nominal absorbed-dose output rate $\dot{D_{o}}=360\frac{Gy}{h}$. The Old method determines that the  wall M1 is not safe because the dose rate ($16.44 \frac{\mu Sv}{week}$) exceeds limit ($10 \frac{\mu Sv}{week}$). However, the new method is saying that is not the case: the correct dose rate is $7,74 \frac{\mu Sv}{week}$ which is less than the limit. The other secondary walls are safe according to the old method, which, as we know, super-dimensionates in this case, then it was not necessary to apply the new method, that is, the leak component $L$ was not measured (NM). In other words, for the cases $U=1 $, we are using the new method to be applied to the hot points of the old method.}
\end{table}

\vspace{1cm}

{\bf B)Situation $U<1$.} Examples

\vspace{1cm}

\underline{\bf BUNKER 5)} \, For a LINAC of 10 MV with primary workload $W=1200\frac{Gy}{week}$ and secondary IMRT leakage load $W_L=3600\frac{Gy}{week}$ and in a nominal absorbed-dose output rate $\dot{D_{o}}=360\frac{Gy}{h}$.

  Control console,  wall C (defined as controlled area) with $T=1$ and $U=\frac{1}{5}$ for scattered component.

We have: 

\be
\frac{1-U}{U} \,.\, \frac{t_L}{t_L-t_{bo}}=\frac{1-\frac{1}{5}}{\frac{1}{5}} \,.\, \frac{10}{10-\frac{10}{3}}=6
\en

Measures with Ionization pressurized chamber Ludlum 9DP serial number 25018216, we obtained for
this wall : 

 \vspace{1cm}

$L_T=6,30 \frac{\mu Sv}{h}$

$L=4,30\frac{\mu Sv}{h}$

$L_{BG}=1,20 \frac{\mu Sv}{h}$

\vspace{0.5cm}

Then:

\be
\frac{Scattered}{Leakage- BG}=\frac{L_T-L}{L- L_{BG}}=0,645
\en

So, because $0,645<6$, it is verified that 

\be\label{b222}
\frac{Scattered}{Leakage- BG} < \frac{1-U}{U} \,.\, \frac{t_L}{t_L-t_{bo}}=6
\en

Then, we are in the {\bf Case B2)} which means $R'-R<0$ i.e. R' is underestimating doses. 

It is obtained from Eq. (\ref{26})

\be
\frac{R'-R}{R}=-0,68
\en

so 

\be
R'= R -0,68\,R = 0,32 \,R 
\en

or equivalently 

\vspace{1cm}

$R'= 10,20 \frac{\mu Sv}{week}$ 

$R = 32,33\frac{\mu Sv}{week}$

\vspace{1cm}

The old method underestimates in $\cong$ 68 $\%$ for de dose rate at the Control Console. 
It is important to note that in this "underestimated" case it is not possible to know in advance how much the method underestimates without measuring the leakage component, that is, without applying the new method. Therefore, in the case $U<1$ and underestimated (case {\bf B2}), it will always be necessary to use the new method because, with the old method, we could be underestimating a dose rate that actually exceeds the allowed limit. This is not the case for this wall because both methods indicates  an adequate dose rate, i.e. , less than the allowed limit for controlled area ($400\frac{\mu Sv}{week}$).
Note that if this region had been hypothetically  considered as uncontrolled area, that is, with allowed limit $L = 20\frac{\mu Sv}{week}$, then the old method would be completely wrong and should be discarded because it would be accepting a shielding that is not really sufficient (the real dose rate being $R = 32,33\frac{\mu Sv}{week}$).  

\begin{figure}[t!]
\begin{center}
\includegraphics[width=12cm]{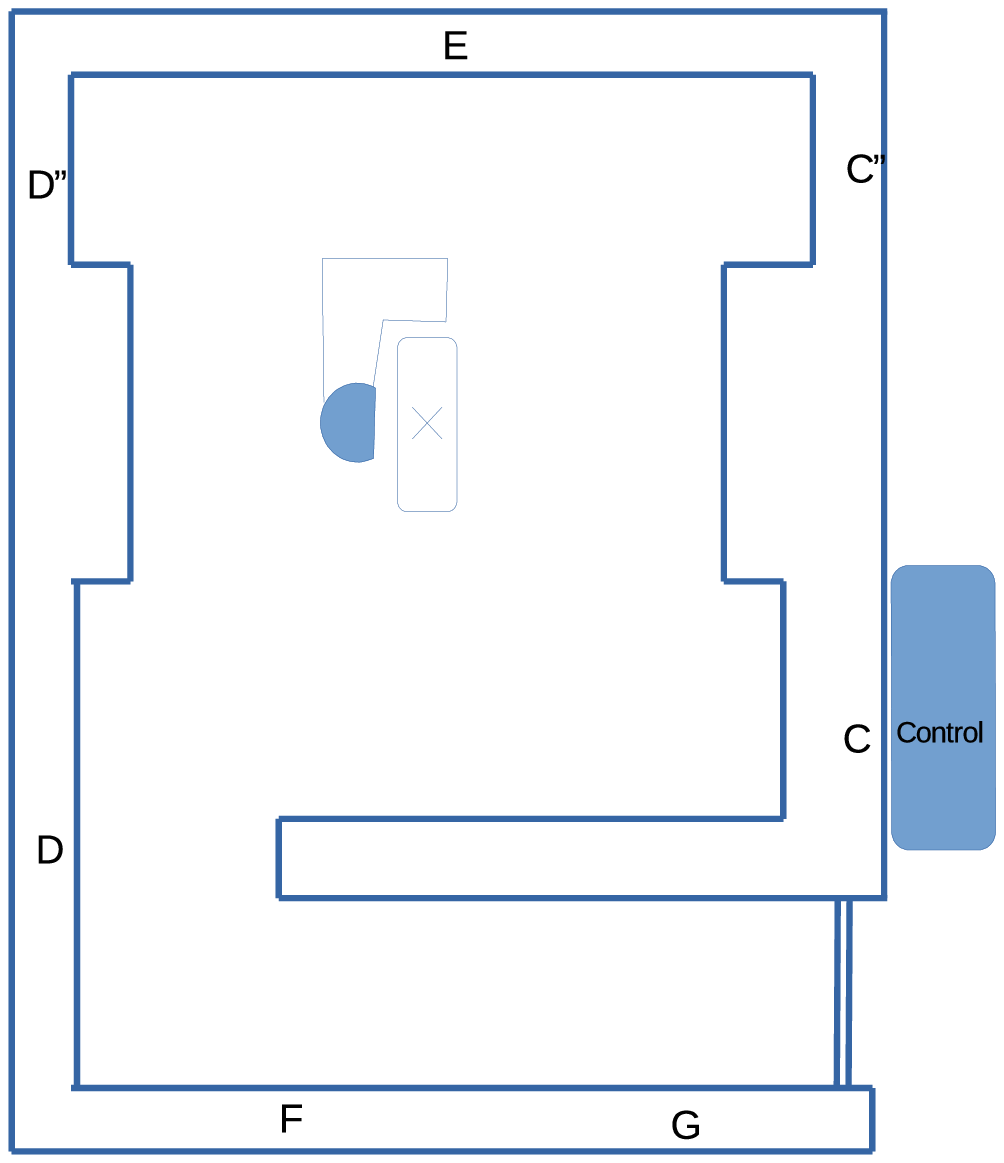}
\caption{Bunker 5. The neighborhoods of each wall are indicated in the table V.}
\label{bunker5}
\end{center}
\end{figure}

For the other secondary walls (Fig. \ref{bunker5} ), measurements and calculations are presented in Table V. We see that on two walls with $U=1$ (the door and TS) the old method, as we already know, gives overestimated values  (situation {\bf A}) or "`pure"'),  but in this case acceptable (less than the allowed limit) and therefore it is enough to measure with the old method. On the other hand, for the walls $C, D, D'$ and $C'$ (all with $U<1$) the old method gives underestimated values (case {\bf B2}), and since we do not know in advance how much, it is necessary to use the new method to verify if the dose rate is acceptable.

\begin{table}[t!]
\begin{tabular}{||c|c|c|c|c|c|c|c|c|c|c||}\hline\hline

\ Wall & Destination & T & U & Area &$L_{T} [\frac{\mu Sv}{h}]$ & $L [\frac{\mu Sv}{h}]$& $L_{BG}[\frac{\mu Sv}{h}]$ &$ R' \frac{\mu Sv}{week} $ & $R\frac{\mu Sv}{week}$ & $ Difference = \frac{R'-R}{R}\times 100$
\\ \hline

P   & Door          &$ \frac{1}{8}$ &$1$& Controlled &$7.00 $ & $3.30 $&$1.20$ & $7.25$ &$ 4.17$ & 74 \%\\ \hline

C   & Control console &  $1$ &$\frac{1}{5}$ & Controlled   &$6.30$ & $4.30$  &$1.20$ & $10.2$ & $32.33$ & -68 \%\\ \hline

D   & External area    & $\frac{1}{8}$ & $\frac{1}{5}$  & Uncontrolled &$2.90 $ & $1.50$ & $1.20$ & $0.17$ & $ 0.49$& -65 \% \\ \hline

D'  & External area    & $\frac{1}{8} $ &$\frac{1}{5}$ & Uncontrolled &$2.20$  & $1.40 $&$1.20$ & $0.10$ &$0.32$  & -69 \%\\ \hline

Ts1   & Technical area  & $\frac{1}{8} $ & $1$ & Uncontrolled&$3.21$  & $1.50$ &$1.20$ & $2.51 $& $1.09$& 130 \%\\ \hline

C'   &    Stabilizer    & $\frac{1}{2} $ &$\frac{1}{5}$ & Uncontrolled  &$5.90$  & $1.36$ &$1.20$ & $1.18 $ & $2.31$ & -49 \%\\ \hline

\hline\hline 
\end{tabular}
\label{tabla5}
\caption{Bunker 5. Measurements for secondaries walls with Ionization pressurized chamber Ludlum 9DP serial number 25018216.  This bunker houses a 10 MV LINAC  with primary workload $W=1200\frac{Gy}{week}$ and secondary IMRT leakage load $W_L=3600\frac{Gy}{week}$ and in a nominal absorbed-dose output rate $\dot{D_{o}}=360\frac{Gy}{h}$. On two walls with $U=1$ (the door P and TS) the old method, as we already know, gives overestimated values  (situation {\bf A}) or "`pure"'),  but in this case acceptable (less than the allowed limit) and therefore it is enough to measure with the old method. On the other hand, for the walls $C, D, D'$ and $C'$ (all with $U<1$) the old method gives underestimated values (case {\bf B2}) and, since we do not know in advance how much, it is necessary to use the new method to verify if the dose rate is acceptable.}
\end{table}

\vspace{1cm}

\underline{\bf BUNKER 6)} \, For a LINAC of 10 MV with primary workload $W=1500\frac{Gy}{week}$ and secondary IMRT leakage load $W_L=4500\frac{Gy}{week}$ and in a nominal absorbed-dose output rate $\dot{D_{o}}=360\frac{Gy}{h}$. We have

\be
t_{bo}=\frac{1500\frac{Gy}{week}}{360\frac{Gy}{h}}= \frac{25}{6}\frac{h}{week} \cong\ 4,17 \frac{h}{week}
\en

\be
t_{L}=\frac{4500\frac{Gy}{week}}{360\frac{Gy}{h}} = 12,5 \frac{h}{week}
\en

Wall {\bf D}, that shields an area partially occupied  by workers (controlled area, allowed limit $P=400\frac{\mu Sv}{week}$), with $T=\frac{1}{2}$ and $U=\frac{1}{5}$ for scattered component. We have: 

\be
\frac{1-U}{U} \,.\, \frac{t_L}{t_L-t_{bo}}=\frac{1-\frac{1}{5}}{\frac{1}{5}} \,.\, \frac{12,5}{12,5-\frac{10}{3}} \cong 5,45
\en

Measures with Ionization pressurized chamber Ludlum 9DP serial number 25009337, we obtained: 

\vspace{1cm}

$L_T=23,20 \frac{\mu Sv}{h}$

$L=0,246\frac{\mu Sv}{h}$

$L_{BG}=0,245 \frac{\mu Sv}{h}$

\vspace{0.5cm}

Then:

\be
\frac{Scattered}{Leakage- BG}=\frac{L_T-L}{L- L_{BG}}=22954
\en

So, because $22954 > 5,45 $, it is verified that 

\be\label{b111}
\frac{Scattered}{Leakage- BG} > \frac{1-U}{U} \,.\, \frac{t_L}{t_L-t_{bo}}=5,45
\en

Then, we are in the {\bf Case B1)} which means $R'-R>0$ i.e. R' is overestimating doses.

It is obtained from Eq. (\ref{26})

\be
\frac{R'-R}{R}=1,998
\en

so 

\be
R'= 1,998 \,R +\,R = 2,998 \,R 
\en

or equivalently 

\vspace{1cm}

$R'= 28,69 \frac{\mu Sv}{week}$ 

$R = 9,57\frac{\mu Sv}{week}$

\vspace{1cm}

The old method overestimates in $\cong$ 200 $\%$ for the dose rate for this area. However, as $R < R'=28,69 \frac{\mu Sv}{week}< 400 \frac{\mu Sv}{week}$,  this method  indicates  an adequate dose rate. But, again, it is important to note that in this "overestimated" case it is not possible to know in advance how much the method overestimates without measuring the leakage component, that is, without applying the new method. Therefore, also in the case $U<1$ and overestimated (case {\bf B1}), it will always be necessary to use the new method because, with the old method, we could be overestimating a dose rate that actually do not exceeds the allowed limit. 
Consider, as for example, that hypothetically this region had been  defined as uncontrolled area , that is, with allowed limit $ P = 20\frac{\mu Sv}{week}$, then the old method would be completely wrong and should be discarded because it would be condemning a shielding that is really safe (the real dose rate being $R = 9,57\frac{\mu Sv}{week}< P $).

\begin{figure}[t!]
\begin{center}
\includegraphics[width=12cm]{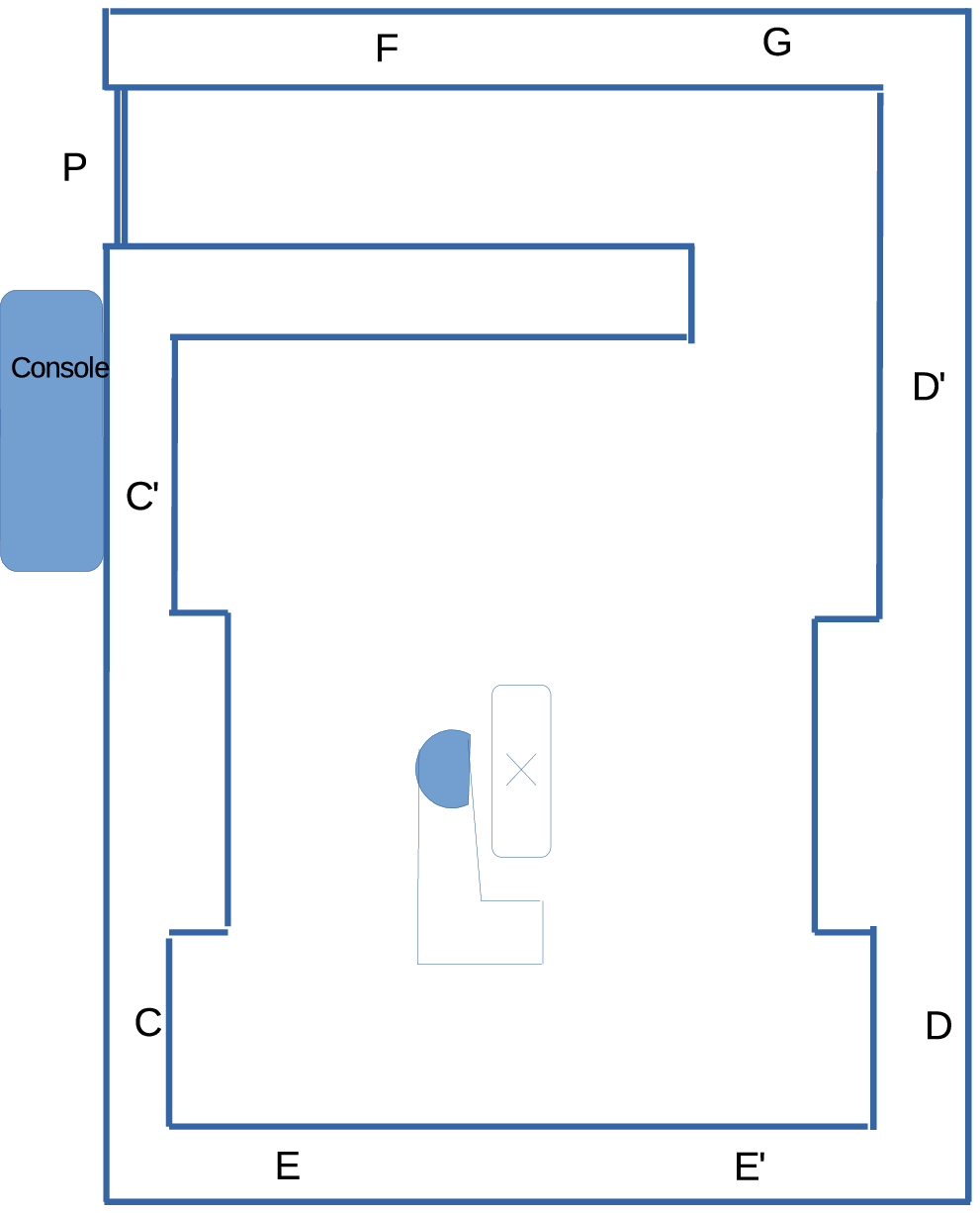} 
\caption{Bunker 6. The neighborhoods of each wall are indicated in the table VI.}
\label{bunker6}
\end{center}
\end{figure}

\begin{table}[t!]
\begin{tabular}{||c|c|c|c|c|c|c|c|c|c|c||}\hline\hline

\ Wall & Destination & T & U & Area &$L_{T} [\frac{\mu Sv}{h}]$ & $L [\frac{\mu Sv}{h}]$& $L_{BG}[\frac{\mu Sv}{h}]$ &$ R' \frac{\mu Sv}{week} $ & $R\frac{\mu Sv}{week}$ & $ Difference = \frac{R'-R}{R}\times 100$
\\ \hline

P   & Door          &$ \frac{1}{4}$ &$1$& Controlled &$12.70 $ & $3.10 $&$0.245$ & $19.461$ &$ 9.461$ & 106 \%\\ \hline

G   & Service &  $\frac{1}{2}$ &$1$ & Uncontrolled   &$0.50$ & $0.26$  &$0.245$ & $1.594$ & $0.594$ & 168 \%\\ \hline

D'   & External area    & $\frac{1}{2}$ & $\frac{1}{5}$  & Uncontrolled &$2.80$  & $0.27 $&$0.245$ & $3.194$ &$1.210$  & 164 \% \\ \hline

D  & External area    & $\frac{1}{2} $ &$\frac{1}{5}$ & Controlled &$23.20$  & $0.246 $&$0.245$ & $28.694$ &$9.570$  & 200 \%\\ \hline

E   & Corridor   & $\frac{1}{2} $ & $1$ & Uncontrolled&$0.80$  & $0.72$ &$0.245$ & $3.469 $& $3.135$ & 11 \%\\ \hline

E'   &    Treatment room   & $\frac{1}{2} $ &$1$ & Controlled  &$0.50$  & $0.44$ &$0.245$ & $3.188 $ & $2.688$ & 19 \%\\ \hline

F   &    External area   & $\frac{1}{2} $ &$1$ & Uncontrolled  &$0.40$  & $0.245$ &$0.245$ & $0.969 $ & $0.323$ & 200 \%\\ \hline

\hline\hline 
\end{tabular}
\label{tabla6}
\caption{Bunker 6. Measurements for secondaries walls with Ionization pressurized chamber Ludlum 9DP serial number 25009337. This bunker houses a LINAC of 10 MV with primary workload $W=1500\frac{Gy}{week}$ and secondary IMRT leakage load $W_L=4500\frac{Gy}{week}$ and in a nominal absorbed-dose output rate $\dot{D_{o}}=360\frac{Gy}{h}$. The old method overestimates in $\cong$ 200 $\%$ for the dose rate for this area. However, as $R < R'=28,69 \frac{\mu Sv}{week}< 400 \frac{\mu Sv}{week}$,  this method  indicates  an adequate dose rate. But, again, it is important to note that in this "overestimated" case it is not possible to know in advance how much the method overestimates without measuring the leakage component, that is, without applying the new method. Therefore, also in the case $U<1$ and overestimated (case {\bf B1}), it will always be necessary to use the new method because, with the old method, we could be overestimating a dose rate that actually do not exceeds the allowed limit. }
\end{table}

For the other secondary walls (Fig. \ref{bunker6} ), measurements and calculations are presented in Table VI. We see that the old method gives overestimated values for all  secondary walls. For the wall $D$ ($U<1$)  we recall that we know this thanks to the new method and we need to apply it in order to decide if the dose rate is acceptable.  For the walls $P$, $G$, $E$, $E'$ and $F$ (all with $U=1$), as we already know from situation {\bf A}, the old method overestimates but we see the values of the  dose rates are still acceptable and both methods allow us to conclude that these walls are safe from the radiological point of view.

\section{Discussion}

We can say that, from the theoretical point of view, the old method for measuring secondary walls is not strictly correct. Based on the analysis and the examples we have presented, we may notice that when $U=1$ the old method overestimates the dose rates but it can still be used to provide an upper bound for dose rates, as long as it does not exceed the allowed limits (legal and project goal).
For the situation $U<1$ , the new method must be used. This makes it possible to decide if the dose rate is adequate, that is, if the shielding is sufficient. The old method does not prove reliable in this case because it is not known if it overestimates or underestimates the dose rates and by how much. To know this, it is necessary to measure the leakage component separately (closing the collimator and the multi-leafs, as we saw in section II) and apply the new method.
One way to work would be the following: situation $U = 1$: the "hot" points obtained with the old method must be verified with the new method.
Situation $U <1$: apply the new method.


\section{Conclusions}

We have compared two methods to perform radiometric surveys for the secondary walls in LINAC radiotherapy services that use IMRT technique. One, the "old" method, employs an adapted formula of the period prior to IMRT technique and is widely used in all radiotherapy services with LINAC. The "new" method uses a formula that is deducted from the theory of structural shielding for IMRT. We found that for secondary walls there are differences: if the secondary wall is "pure" ($U = 1$) the old method always super-estimate the dose rates. If the secondary wall is "not pure" ($U <1$) the old method can both overestimate or underestimate the dose rates. We have carried out a series of measurements that  verify these conclusions. An optimized procedure is proposed: in the case of "pure" secondary walls ($U = 1$) measure first with the old method and, at hot points, discard it and use the new method; in the case of secondary walls with $U <1$ only the new method must be used.


\section*{Acknowledgments}

We thank CNEN/CGMI / MCTI-Brazil for technical support and to all the country's radiotherapy services that kindly have allowed us to make the measurements.



\begin{thebibliography}{99}


\bibitem{paiva}{E. de Paiva, R. A. Giannoni, A. F. Velasco, R. R. A. Brito, A. C. M. Dovales, L. V. de Sá, L. A. R. da Rosa, Radiometric survey of teletherapy treatment rooms in Brazil, Radiation Protection Dosimetry, Volume 138, Issue 4, March 2010, Pages 402–406, https://doi.org/10.1093/rpd/ncp264}
\bibitem{NCRP151}{Structural Shielding design and evaluation for megavoltagem X and Gamma- Ray radiotherapy Facilities. NCRP report \#151 (2005).}
\bibitem{kairn}{Kairn T, Crowe SB, Trapp JV. Correcting radiation survey data to account for increased leakage during intensity modulated radiotherapy treatments. Med Phys. 2013 Nov;40(11):111708. doi: 10.1118/1.4823776. PMID: 24320416.}
\bibitem{McGinley}{Patton H. McGinley, Shielding Techniques for Radiation Oncology Facilities, Medical Physics Publishing, (1998).}
\bibitem{apostila}{Eug\^enio Del Vigna e Rossana C. Falc\~ao, Blindagem em Radioterapia, T\'ecnicas e Normas, Eug\^enio PQRT- INCA (2000).} 
\bibitem{preprint} E. S. Santini, R.V.de Oliveira, N. Couto ,  C. Salata, P.A.P. Leal, F.C. da S. Teixeira, G.S. Joana, arXiv:2112.14230.



\end{thebibliography}
\end{document}